\documentstyle[12pt]{article}
\newcommand{\be}{\begin{eqnarray}}
\newcommand{\ee}{\end{eqnarray}}
\newcommand{\ba}{\begin{array}}
\newcommand{\ea}{\end{array}}

\def\o{\over}
\def\l{\lambda}
\def\b{\beta}
\def\bl{\b_\l}
\def\bm{\b_M}
\def\f{\varphi}
\def\G{\Gamma}
\def\gf{\gamma_\f}
\def\fb{\bar\f}
\def\fm{\bar\f_0}
\def\p{\partial}
\def\Zf{Z_\f}
\def\io{\hbox{$\bigcirc\kern-2.3pt{\bf\cdot}$\kern 2.3pt}}
\def\bub{\hbox{${\bf\cdot}\kern-5pt\bigcirc\kern-5pt{\bf\cdot}$}}     
\def\fe{\hbox{$\bigcirc$}}   
\def\itro{\hbox{${\bf\cdot}\kern-5pt\bigcirc\kern-5pt{\bf :}$}}
\begin{document}
\begin{flushright}
DIAS-STP-95-37\\
\today
\end{flushright}
\vskip 22 truept
\begin{center}
{\LARGE Finite-Temperature Renormalization Group}

\vspace{.2cm}

{\LARGE Predictions: The Critical Temperature,} 

\vspace{.2cm}

{\LARGE Exponents and  Amplitude Ratios}

\vspace{1.cm}

{\bf F.\ Freire}\footnote{Institut f\"ur Theoretische Physik,
Universit\"at Heidelberg,
Philosophenweg 16, 69120 Heidelberg, Germany},
{\bf Denjoe\ O' Connor}\footnote{School of Theoretical Physics,
Dublin Institute for Advanced Studies, 10 Burlington Road, Dublin 4,
Ireland.},
{\bf C.R.\ Stephens}\footnote{Instituto de Ciencias Nucleares,
U.N.A.M., A. Postal 70-543, 04510 Mexico D.F., Mexico.}.
{\bf and M.A.\ van Eijck}\footnote{Institute for Theoretical Physics,
University of Amsterdam,
Valckenierstraat 65, NL-1018 XE Amsterdam, Netherlands}\end{center}
\vskip 1.3truein
{\bf Abstract:}
$\l\f^4$ theory at finite temperature suffers from infrared divergences 
near the temperature at which the symmetry is restored. 
These divergences are handled using renormalization group methods.
Flow equations which use a fiducial mass as flow parameter
are well adapted to predicting the non-trivial critical exponents whose 
presence is reflected in these divergences. 
Using a fiducial temperature as flow parameter,
we predict the critical temperature, at which the mass vanishes,
in terms of  the zero-temperature mass and coupling.
We find some universal amplitude
ratios which connect the broken and symmetric phases of the theory
which agree well with those of the
three-dimensional Ising model obtained from
high- and low-temperature series expansions.
\vfill\eject

\section{Introduction}
The problem of ultraviolet (UV) divergences is well known to have its solution 
in the proper renormalization of a theory. What is not so well 
appreciated is that infrared (IR) divergences can equally well be treated 
by such methods. In finite temperature field theory UV divergences are 
temperature independent while the IR divergences are temperature dependent. 
Thus for the satisfactory removal of these finite temperature divergences 
it is necessary to have a subtraction scheme
which is also temperature dependent. When this subtraction scheme is 
associated with a multiplicative renormalization group then the flow equations
of that group will of necessity be temperature dependent. This is a special case
of a more general situation where IR singularities depend on 
``environmental'' parameters. In such settings the flow equations of 
a well adapted renormalization group (RG) should also depend on those 
parameters. Such renormalization groups we call ``environmentally 
friendly'' renormalization groups. 

The techniques of ``environmentally friendly renormalization''
\cite{NucJphysa,EnvfRG} offer a quite general approach to
investigating, both qualitatively and quantitatively, the change from 
one type of critical behaviour to another as an ``environmental'' 
parameter is varied. Such a change of critical behaviour is known, 
in the terminology of critical phenomena, as a crossover.
In the context of finite-temperature field theory, a temperature dependent
reparametrization defined by normalization conditions provides a method
of tracking the evolution of the effective degrees of freedom
as a function of both scale and temperature.
The resulting finite-temperature renormalization group \cite{FTRG}
is therefore, in principle, environmentally friendly.

The problem of finite temperature field theory in its Euclidean formulation
is exactly that of critical phenomena on the manifold $R^3\times S^1$.
In this setting it falls under the heading of 
critical phenomena with finite size corrections. The finite size of the 
$S^1$ is $L=1/T$ (more conventionally
one uses the symbol $\beta=1/T$).    
In the series of papers \cite{NucJphysa,EnvfRG,prevfintemp,physreve}
a one-parameter family of reparametrizations at
fixed temperature $T$, parametrized by an arbitrary fiducial
finite-temperature mass, was considered. 
This rendered the complete crossover
for all values of $T$ accessible in one uniform
perturbation expansion. The crossover was analyzed to two loops
and the phase transition shown to be second order, characterized by
three-dimensional critical exponents.
More recently \cite{ampratios} we found reliable predictions, for the 
critical regime in terms of the zero-temperature parameters of the theory, 
by considering a 
one parameter family of renormalized couplings
parametrized by an arbitrary fiducial temperature $\tau$.

In the neighbourhood of the critical point where the symmetry is 
restored the mass and coupling vanish continuously at $T_c$ and behave
as ${(f_1^{\pm})}^{-1}|T-T_c|^{\nu}$ and $l^{\pm}|T-T_c|^{\nu(4-d_c)}$
respectively,
as expected from critical phenomena (and in agreement with \cite{NucJphysa}).
Here, $d_c$ is the reduced dimension in the critical
regime, $\nu$ and $\eta$ are characteristic exponents, 
$f_1^{\pm}$ and $l^{\pm}$ are amplitudes, the $\pm$ referring to above 
and below the critical temperature respectively. 
For $\lambda\phi^4$ theory and the physical dimension 
$d=4$, one finds $d_c=3$ and our two loop results gave 
$\nu=0.639$ and $\eta=0.0329$ \cite{EnvfRG} while the best available 
estimates for these exponents are $\nu=0.6310\pm0.0015$ and 
$\eta=0.0375\pm0.0025$ \cite{ZinnJustin}. 

Amplitudes differ above and below the critical point, however certain ratios 
of these amplitudes, like critical exponents, are universal numbers.
In the present context this means they are independent of the zero-temperature
mass and coupling which we use to parametrize the theory.
We found \cite{ampratios} that the amplitude ratio 
$f_1^{+}/f_1^{-}=1.92$ which is in good agreement
with the best series expansion results of Liu and Fisher \cite{LiuFisher}
who obtain $1.96\pm 0.01$ (this should be compared with the result $1.41$
obtained from a tree level analysis where fluctuations are ignored
and $1.91$ from the $\epsilon$ expansion at order $\epsilon^2$ \cite{BGZJ} 
for $\epsilon=1$). The other principal amplitude ratio we studied was that
associated with the two point vertex function $\Gamma^{(2)}$ at zero 
momentum which we denote by $M^2$. Near $T_c$ it takes the form
$M^2_\pm=C^\pm\vert T-T_c\vert^{\gamma}$ and we found that
$C^+/C^-=4$ while the best estimates of Liu and Fisher from 
series expansions give $C^+/C^-=4.95\pm0.15$.

\section{Finite-temperature renormalization}
We consider a $\f^4$ theory in equilibrium with a 
thermal bath at temperature $T=1/\beta$. It is described by the 
Euclidean action 
\be
S[\f_{B}]=\int_0^\b dt\int
d^{d-1}x\left[{1\o2}(\nabla\f_{B})^2+{1\o2}M^2_{B}\f_{B}^2+
{\l_{B}\o4!}\f_{B}^{4}\right].\label{action}
\ee
The effective potential $V$, or effective action
($\G$) per unit volume is a function of the renormalized field expectation
value $\fb=\Zf^{-1/2}\fb_B$ and provides a convenient summary of many of the
thermodynamic properties of the model. To access an arbitrary value of 
$\fb$ it is necessary to couple in a constant external current $J$, 
which will produce this expectation value. In fact the effective potential is
a function of $\fb^2$ and the relation between $\fb$ and $J$ is given by the 
equation of state
\be
\G^{(1)}=\G^{(2)}_t\fb=J,\label{eqnofstate}
\ee
which also serves to define $\G^{(2)}_t$ again as a function of $\fb^2$.
We further define $\G^{(4)}_t$ through
\be
\G^{(2)}=\G^{(2)}_t+{\G^{(4)}_t\o3}\fb^2,\label{dfnofgft}
\ee
and more generally each $n$-point vertex function
admits a decomposition in terms of vertex functions $\G^{(n)}_t$ 
which are functions of $\fb^2$.

For $J=0$ there are two possible solutions of (\ref{eqnofstate}): Either
$\fb=0$ with $\G^{(2)}_t\neq0$ or $\fb=\fm\neq0$ and $\G^{(2)}_t=0$. 
For fixed temperature it is natural to consider this model and hence the 
effective potential as a function of both $\fb$ and $M_B^2$. In the entire
$(M_B^2,\fb)$ plane both branches of (\ref{eqnofstate}) are realized.
If we locate our origin at the point where the model switches 
from one branch to another, then we can describe the plane in terms
of new co-ordinates $(t_B,\fb)$ where $t_B=M_B^2+\Delta(T)$.
It turns out that it is necessary to multiplicatively renormalize $t_B$ 
in a similar fashion to $\fb$ by defining  $t=Z_{\varphi^2}^{-1}t_B$. 
In fact to control the infrared divergences that necessarily arise close to the 
origin of the $(t_B,\fb)$ plane, which corresponds to the location of a 
second order phase transition, it is necessary to have both 
$Z_{\varphi}$ and $Z_{\varphi^2}$ temperature dependent. If one changes 
$T$ then the picture remains qualitatively the
same but the origin and detailed shape of the 
co-existence curve change somewhat.

We define the $Z$ factors above by the following
normalization conditions
\be
\left.{\p\o\p 
p^2}\G^{(2)}_t(p,\fb(\tau,\kappa),t(\tau,\kappa),\l(\tau,\kappa),\tau)\right|_{p=0}
&=&1,\\
{\G^{(2)}(0,\fb(\tau,\kappa),t(\kappa,\tau),\l(\tau,\kappa),\tau)\o 
1+\left.{\fb^2\o3}{\p\o\p p^2}\G^{(4)}_t(\tau,\kappa)\right|_{p=0}}
&=&\kappa^2(\tau),\label{massCnd}\\
\G^{(2,1)}_t(0,\fb(\tau,\kappa),t(\tau,\kappa),\l(\tau,\kappa),\tau)&=&1,\label{tcnd}\\
\G^{(4)}_t(0,\fb(\tau,\kappa),t(\tau,\kappa),\l(\tau,\kappa),\tau)&=&\l(\tau,\kappa).\label{normcnds}
\ee
These normalization conditions allow us to present the two different 
schemes that we have used. The first corresponds to flowing $\kappa$ 
while $\tau$ is set to the physical temperature $T$. It is convenient to define
$h=4\lambda\kappa^2\itro$ and $z={\kappa\over T}$ 
in which case the flow equations become
\be
\kappa{d h\o d\kappa}&=&\beta(h,z),\\
\kappa{d\ln Z^{-1}_{\varphi^2}\o d\kappa}&=&\gamma_{\varphi^2}(h,z),\\
\kappa{d\ln Z_{\varphi}\o d\kappa}&=&\gamma_{\varphi}(h,z).
\ee
In \cite{EnvfRG} the functions $\beta(h,z)$, $\gamma_{\varphi^2}(h,z)$ and
$\gamma_{\varphi}(h,z)$ were calculated 
to two loops. Then using  Pad\'e resummation techniques the 
resulting equations were solved to find quite accurate expressions
for the resulting Wilson functions. These capture the dominant
part of the critical behaviour over the entire range of 
temperatures and especially in the neighbourhood of the critical 
temperature. 
For example the relationship between the inverse correlation length 
and the parameter $t$ that tells us the deviation from the critical 
region is given by 
\be
t(m,\kappa)=\kappa^2\int_0^{m}{dx\o x}(2-\gamma_{\varphi})e^{\int_\kappa^x(2-\gamma_{\varphi^2}){dy\over y}} .
\ee
The principal quantity that is missed in this approach is the critical 
temperature or equivalently the quantity $\Delta(T)$ described above.

The second prescription involves flowing $\tau$ at fixed $M_B$
and allows us to predict the critical temperature in terms of the zero 
temperature parameters, precisely as one would wish to do in 
the case of the standard model.
In this prescription we choose to follow the flow of 
$M^2=\left.\Gamma^{(2)}\right|_{p=0}$. 
The differential equations which describe an infinitesimal change in
normalization point with fixed bare parameters are
\be
\tau{d\ln Z_\f(\tau)\o d\tau}=\gf,\qquad
\tau{d\ln Z_{\varphi^2}^{-1} \over d\tau}=\gamma_{\varphi^2},\qquad
\tau{dM^2(\tau)\o d\tau}=\bm,\qquad
\tau{d\l(\tau)\o d\tau}=\bl.
\ee
For $J=0$ the flow functions $\bm$, $\bl$ and $\gf$ take different
functional forms above and below the branching point which corresponds to 
$T=T_c$, the critical temperature of the model.
The flow functions to one loop, are given by
\be
\gf & = & 0, \\
\gamma_{\varphi^2}& = &-{1\over2}\l\tau{d\over d\tau}\bub , \\
\bm & = &\left\{\ba{ll}{\l\o 2} \tau{\p\io\o\p\tau},& \tau>T_c\\
& \\
-\l \left(\tau{\p\io\o\p\tau}+{3\o2}M^2\tau{\p\io\o\p\tau}\right),
&\tau<T_c, \ea\right.\\
\bl & = &-{3\o2}\l^2\tau{d\o d\tau}\bub .\label{couplingflow}
\ee
The symbol $\fe$ with $k$ dots stands for the one-loop diagram with 
$k$ propagators, without vertex factors, and zero external momentum.

After solving the flow equations we are free to choose the reference
temperature $\tau$ equal to the actual temperature $T$ of interest.
In fact this is essential if one wishes to obtain perturbatively sensible
results for physical quantities \cite{Filipe}.
Because of the renormalization conditions (\ref{massCnd}, \ref{normcnds})
the parameters $M(T)$ and $\lambda(T)$ therefore describe the behaviour
of the vertex functions $\G^{(2)}$ and $\G^{(4)}_t$ at zero momentum.
With these equations one is also able to determine the critical temperature
in terms of the zero-temperature parameters $m(0)$ and $\l(0)$.
As may be expected on dimensional grounds $T_c$ is proportional to $m(0)$,
the constant of proportionality being a function of $\l(0)$.
Our critical temperature is larger than that found by others 
\cite{DolanJackiw,Lawrie,TdrWett} and the value $t=\sqrt{12}$ is 
approached in the zero-coupling limit.
Our results for the critical temperature are most closely comparable 
with ``coarse-graining'' type renormalization groups such as used in 
\cite{TdrWett}. They also found the critical
temperature increases as a function of $\lambda(0)$ but at a
rate substantially less than we find. Also their results are dependent on
the particular type of cutoff function used.

If we take values for $\l(0)$ and $M(0)$ to be those associated with
estimates for the equivalent parameters in the Higgs sector of the
standard model \cite{FordJonesetc} we find, for $\l(0)=1.98$ and $M(0)=200$GeV
that $T_c=613$GeV. By comparison Dolan and Jackiw's result 
\cite{DolanJackiw} gives $492.4$GeV. For $\l(0)=3.00$ and $M(0)=246$GeV
the corresponding results are $T_c=639$GeV and $492.0$GeV respectively.

Since $M(T)\ll T$ near $T_c$ the finite-temperature four-dimensional theory
reduces there to a three-dimensional Landau-Ginzburg model,
in accord with finite size scaling (see \cite{Barber} for a review). In 
the neighbourhood or the critical temperature the general vertex functions 
have the form
\be
\G^{(n)}_{\pm}=\gamma^{(n)}_{\pm}|T-T_c|^{\nu\left(d_c-n{d_c-2+\eta\o2}\right)},
\label{scaling}
\ee
where $d_c=d-1$ is the reduced dimension at the 
critical point, $\nu$ and $\eta$ are 
critical exponents and $\gamma^{(n)}_{\pm}$ are amplitudes.
The appearance of the critical exponent $\nu$ is unusual in particle physics.
It reflects the need for composite operator renormalization 
and the physical dependence of $m(T)$ on temperature. This ensures that
the exponent $\nu$ is physically accessible in finite temperature 
field theory whereas it is usually not experimentally observable in a 
particle physics context as the dependence of the renormalized 
mass on the bare mass is experimentally inaccessible.
As we see here, however, the exponent $\nu$ in fact plays a highly 
significant role near the critical point.

{}From the solutions (\ref{couplingflow})
by noting that the function $\bub$ behaves like $T/8\pi M$
we see that as the critical point is approached,
the coupling vanishes
\cite{NucJphysa,EnvfRG,prevfintemp,TdrWett}.
Hence, near the critical temperature we have 
$M^2_{\pm}={(C^{\pm})}^{-1}|T-T_c|^{\gamma}$ with
$\l_{\pm}=l^{\pm}|T-T_c|^{\nu}$ and $m_{\pm}={(f_1^{\pm})}^{-1}|T-T_c|^{\nu}$,
where we use the notation of Liu and Fisher \cite{LiuFisher} for the amplitudes.
and $\gamma=\nu(2-\eta)$. Our temperature flow equations 
give the one loop values for the exponents 
$\nu=1$ and $\eta=0$, which are not as good as those obtained 
by flowing the mass parameter where our two loop Pad\'e results 
\cite{EnvfRG} for these exponents are $\nu=0.639$ and $\eta=0.0329$. 
We expect, however, that the agreement
between the two schemes should improve at higher orders. 
The exponents $\nu$ and $\eta$ have been estimated by different methods 
(see \cite{ZinnJustin}) with the results $\nu=0.6310\pm 0.0015$ 
and  $\eta=0.0375\pm 0.0025$. 

The temperature flow 
scheme gives the amplitude ratios
\be
{C^+\o C^-}=4,\qquad\qquad
{f_1^+\o f_1^-}=2\sqrt{12\o13}\approx1.92,\qquad\qquad
{l^+\o l^-}={1\o2}.
\ee
The fact that these are not one is indicative of a 
cusp in the graphs of mass and coupling versus temperature as the theory 
passes through the critical temperature.
These ratios are universal numbers analogous to the critical exponents.
The best estimates for the amplitude ratios are the high- and low-temperature
series expansion results of Liu and Fisher \cite{LiuFisher} who find
\be
{C^+\o C^-}=4.95\pm0.15,\qquad\qquad
{f_1^+\o f_1^-}=1.96\pm0.01,\qquad\qquad
\ee
which our results are in good agreement with. By comparison: 
at tree level  (mean field theory) $C^+/C^-=2$ and $f_1^+/f_1^-=1.41$, 
whilst in the $\epsilon$ expansion at order $\epsilon^2$, 
assuming dimensional reduction, 
$C^+/C^-=4.8$ and $f_1^+/f_1^-=1.91$ \cite{BGZJ}.

We see here that the amplitude ratios are substantially better than the 
results for critical exponents. This indicates a complimentarity
between the current approach of flowing the environment, temperature, and 
that of \cite{NucJphysa,EnvfRG} where the flow parameter was the finite 
temperature mass.  At one loop the latter group gives better results for 
amplitudes whereas the former gives better results 
for exponents, but both schemes should converge to the same results 
as one goes to higher orders.

\end{document}